\RequirePackage[mathlines]{lineno} 
\documentclass[a4paper,11pt]{article}
\pdfoutput=1 

\usepackage{jheppub} 

\usepackage[T1]{fontenc} 

\usepackage{threeparttable}
\usepackage{multirow}
\usepackage{subfigure}
\usepackage{float}
\usepackage{adjustbox}
\usepackage{tikz-feynman}
\usepackage{diagbox}
\let\oldequation\equation
\let\oldendequation\endequation
\renewenvironment{equation}
  {\linenomathNonumbers\oldequation}
  {\oldendequation\endlinenomath}

\def \ee   {e^+e^-}

\def \piz  {\pi^0}
\def \pip  {\pi^+}
\def \pim  {\pi^-}

\def \gev  {\mbox{GeV}}

\def \mev  {\mbox{MeV}}

\begin{document}
\setrunninglinenumbers

\title{\boldmath Measurements of branching fractions of $\Lambda_{c}^{+}\to\Sigma^{0}K_{S}^{0}\pi^{+}$ and $\Lambda_{c}^{+}\to\Sigma^{0}K_{S}^{0}K^{+}$}

\collaborationImg{\includegraphics[height=30mm,angle=90]{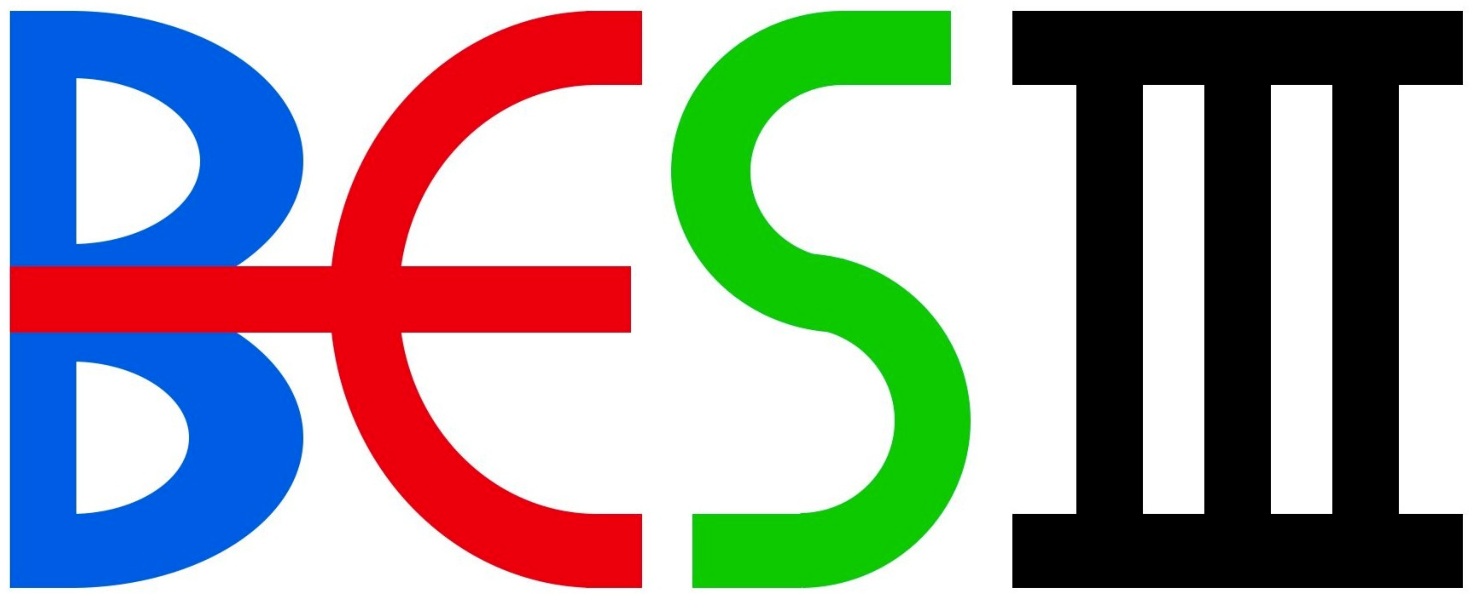}}
\collaboration{The BESIII collaboration}
\emailAdd{BESIII-publications@ihep.ac.cn}

\abstract{
 Based on a data sample corresponding to an integrated luminosity of
 6.4~fb$^{-1}$ of $e^+e^-$ annihilation and collected with the BESIII
 detector at 13 center-of-mass energy points ranging between 4.600~GeV
 and 4.950~GeV, we report the first observation of the singly
 Cabibbo-suppressed decay $\Lambda_c^+ \to \Sigma^0 K_S^0 \pi^+$ with
 a statistical significance of 5.9$\sigma$. The branching fraction is
 determined to be $\mathcal{B}(\Lambda_c^+ \to \Sigma^0 K_S^0 \pi^+) =
 (0.58 \pm 0.14_{\rm stat.} \pm 0.04_{\rm syst.}) \times 10^{-3}$.  In
 addition, the decay $\Lambda_c^+\to\Sigma^0K_{S}^{0} K^+$ has also
 been investigated, and the evidence for this decay is obtained
 with a statistical significance of 3.7$\sigma$. 
}


\newcommand{\BESIIIorcid}[1]{\href{https://orcid.org/#1}{\hspace*{0.1em}\raisebox{-0.45ex}{\includegraphics[width=1em]{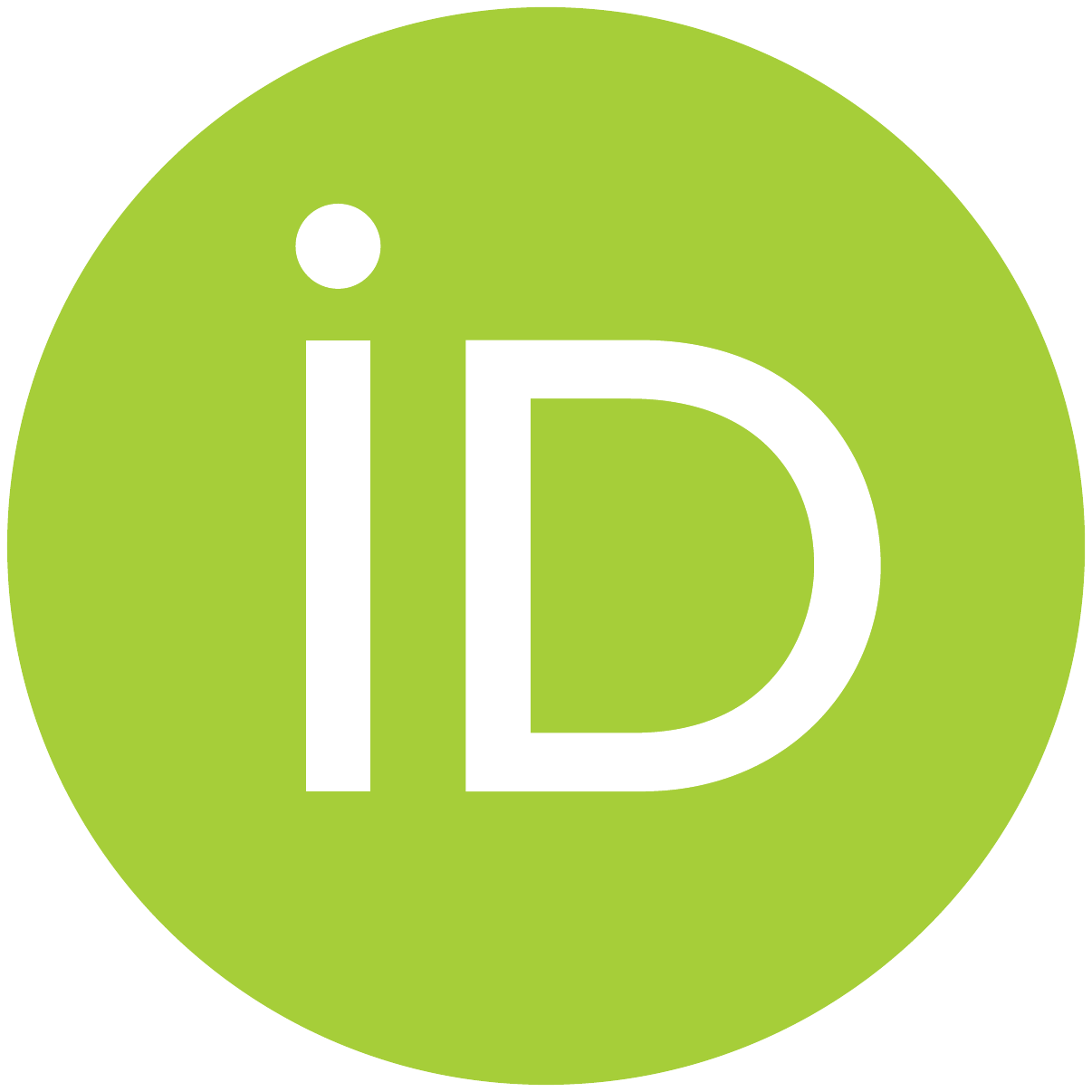}}}}
\maketitle
\flushbottom

\section{INTRODUCTION} \label{sec:introduction} \hspace{1.5em} The
study of the dynamics of charmed baryon decays is crucial to elucidate
the weak and strong interactions in the Standard Model (SM) of
particle physics~\cite{Politzer:1973QCD1,Gross:1973QCD2}. The ground
state of the singly-charmed baryon $\Lambda_c^+$ with a spin-parity
$J^{P}={\frac{1}{2}}^+$~\cite{BESIII:2020LambdacSP} was first observed
in the 1980s~\cite{Abrams:1979FirstLambdac}. Since 2014, there has
been notable progress on the weak hadronic decays of $\Lambda_c^+$,
$\Xi_c^{+(0)}$, and $\Omega_c^0$, both experimentally and
theoretically
~\cite{Cheng:2015iom,Cheng:2021qpd,Li:2021iwf,Li:2025nzx,Wang:2024wrm}. This
has provided crucial information about the properties of all the
singly-charmed baryons and the searches for doubly-charmed baryons
($\Xi_{cc}$ and $\Omega_{cc}$)~\cite{Yu:2017zst}. However, our
understanding of the decay dynamics of charmed baryons is still
limited, due to the lack of high-precision experimental measurements
and the difficulties in the theoretical treatment of strong
interaction effects.

For charmed meson decays, both the annihilation and exchange mechanisms
are either helicity- or color-suppressed, so the spectator diagram is
considered to be the dominant mechanism~\cite{Uppal:1994pt}, and the
factorization approach~\cite{Tetlalmatzi-Xolocotzi:2024ztd}
has been successfully applied to charmed meson decays.  In contrast,
for charmed baryons, there is no helicity and color suppression since
there is an additional light quark, so the W-exchange contributions
may be large, and the factorization approximation generally does not
work. Consequently‌, understanding W-exchange contributions is of
importance for the description of non-leptonic charmed
baryon decay.

To overcome these challenges~\cite{Zenczykowski:1993hw}, significant
efforts have been devoted to developing alternative theoretical
frameworks for charmed hadron
decays~\cite{Cheng:1991sn,Cheng:2018hwl}. These approaches recognize
the necessity of considering nonfactorizable effects.  The $SU(3)$
flavor symmetry ($SU(3)_F$) method has been tested as a useful tool in
both the beauty and charmed hadron decays~\cite{Savage:1989qr,
Geng:2017mxn}. Its feasibility has been established in two-body and
three-body semileptonic charmed baryon weak decays.  The contributions
of the color-antisymmetric and -symmetric part of the effective
Hamiltonian have been considered and fit to experimental data points
to predict the branching fractions (BF) of unknown
decays~\cite{Geng:2024sgq}. Experimental investigations are of great
importance in validating the theoretical methods.

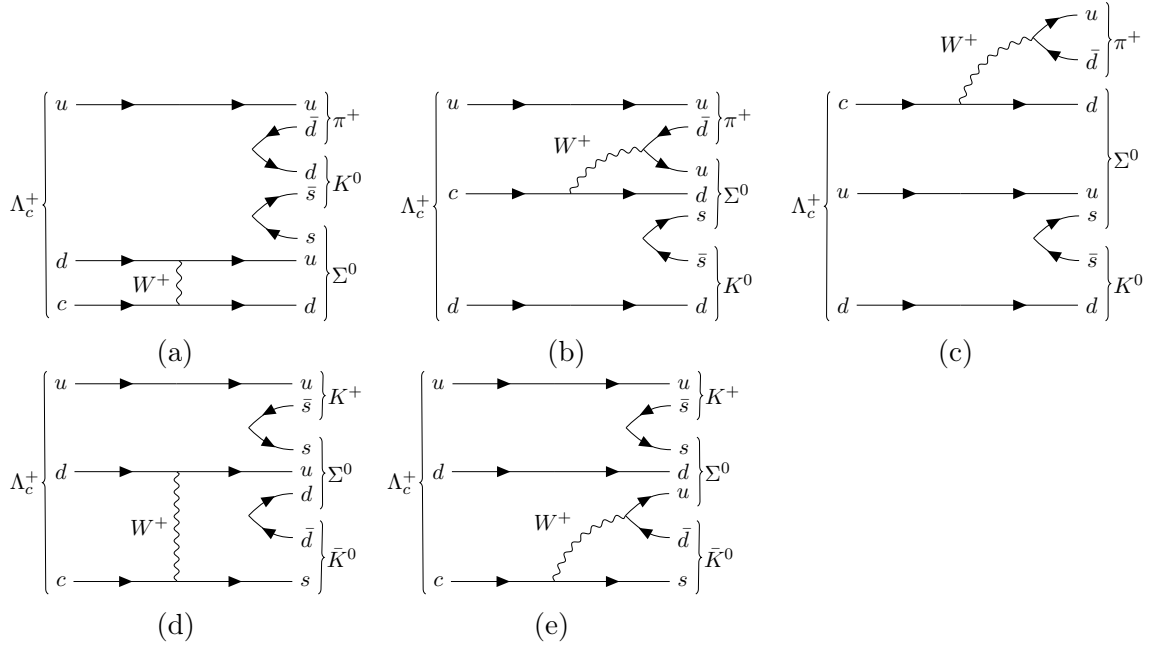
\begin{figure}[h!]
\begin{minipage}[b]{.32\linewidth}
  \begin{adjustbox}{width=\linewidth}
\begin{tikzpicture}
  \begin{feynman}

    \vertex (a1) {\( u\)};
    \vertex[right=2cm of a1] (a2);
    \vertex[right=2cm of a2] (a3){\(u\)};

    \vertex[below=1em of a3] (c3) {\(\bar d\)};
    \vertex[below=2em of c3] (c1) {\( d\)};
    \vertex at ($(c1)!0.5!(c3) - (1cm, 0)$) (c2);
    
    \vertex[below=3em of c3] (d1) {\(\bar s\)};
    \vertex[below=2em of d1] (d3) {\(s\)};
    \vertex at ($(d1)!0.5!(d3) - (1cm, 0)$) (d2);
    
           \vertex[below=7em of a1] (b1) {\(d\)};
    \vertex[right=2cm of b1] (b2);
    \vertex[right=2cm of b2] (b3) {\(u\)};
    
    \vertex[below=2em of b1] (e1) {\(c\)};
    \vertex[right=2cm of e1] (e2);
    \vertex[right=2cm of e2] (e3) {\(d\)};

    \diagram* {
      {[edges=fermion]
        (a1) -- (a2) -- (a3),
        (b1) -- (b2) -- (b3),
        (e1) -- (e2) -- (e3),
      },
    (e2) -- [boson,right, edge label=\(W^+\)] (b2),

      (c3) -- [fermion, out=-180, in=45] (c2) -- [fermion, out=-45, in=180] (c1),
      (d3) -- [fermion, out  =180, in=-45] (d2) -- [fermion, out=45, in=180] (d1),
    };

    \draw [decoration={brace}, decorate] (e1.south west) -- (a1.north west)
          node [pos=0.5, left] {\(\Lambda_{c}^{+}\)};
    \draw [decoration={brace}, decorate] (c1.north east) -- (d1.south east)
          node [pos=0.5, right] {\( K^{0}\)};
    \draw [decoration={brace}, decorate] (a3.north east) -- (c3.south east)
         node [pos=0.5, right] {\(\pi^{+}\)};
    \draw [decoration={brace}, decorate] (d3.north east) -- (e3.south east)
         node [pos=0.5, right] {\(\Sigma^{0}\)};
  
  \end{feynman}
  \end{tikzpicture}
    \end{adjustbox}
   \put(-80,-15){(a)}
\end{minipage}
\begin{minipage}[b]{.32\linewidth}
  \begin{adjustbox}{width=\linewidth}
\begin{tikzpicture}

  \begin{feynman}

    \vertex (a1) {\( u\)};
    \vertex[right=2cm of a1] (a2);
    \vertex[right=2cm of a2] (a3){\(u\)};

    \vertex[below=4em of a1] (e1) {\(c\)};
    \vertex[right=2cm of e1] (e2);
    \vertex[right=2cm of e2] (e3) {\(d\)};

    \vertex[below=5em of e1] (b1) {\(d\)};
    \vertex[right=2cm of b1] (b2);
    \vertex[right=2cm of b2] (b3) {\(d\)};

    \vertex[below=1em of a3] (c3) {\(\bar d\)};
    \vertex[below=2em of c3] (c1) {\( u\)};
    \vertex at ($(c1)!0.5!(c3) - (1cm, 0)$) (c2);
    
    \vertex[below=1em of e3] (d1) {\( s\)};
    \vertex[below=2em of d1] (d3) {\(\bar s\)};
    \vertex at ($(d1)!0.5!(d3) - (1cm, 0)$) (d2);

    \diagram* {
      {[edges=fermion]
        (a1) -- (a2) -- (a3),
        (b1) -- (b2) -- (b3),
        (e1) -- (e2) -- (e3),
      },
      (e2) -- [boson,bend left, edge label=\(W^+\)] (c2),

      (c3) -- [fermion, out=-180, in=45] (c2) -- [fermion, out=-45, in=180] (c1),
      (d3) -- [fermion, out=180, in=-45] (d2) -- [fermion, out=45, in=180] (d1),
    };

    \draw [decoration={brace}, decorate] (b1.south west) -- (a1.north west)
          node [pos=0.5, left] {\(\Lambda_{c}^{+}\)};
    \draw [decoration={brace}, decorate] (c1.north east) -- (d1.south east)
          node [pos=0.5, right] {\(\Sigma^{0}\)};
    \draw [decoration={brace}, decorate] (a3.north east) -- (c3.south east)
         node [pos=0.5, right] {\(\pi^{+}\)};
    \draw [decoration={brace}, decorate] (d3.north east) -- (b3.south east)
         node [pos=0.5, right] {\( K^{0}\)};
  \end{feynman}
  \end{tikzpicture}
\end{adjustbox}
\put(-80,-15){(b)}
\end{minipage}
\begin{minipage}[b]{.32\linewidth}
  \begin{adjustbox}{width=\linewidth}
\begin{tikzpicture}
  \begin{feynman}

    \vertex (a1) {\( c\)};
    \vertex[right=2cm of a1] (a2);
    \vertex[right=2cm of a2] (a3){\(d\)};
    
    \vertex[below=4em of a1] (b1) {\(u\)};
    \vertex[right=2cm of b1] (b2);
    \vertex[right=2cm of b2] (b3) {\(u\)};

    \vertex[below=1em of b3] (c3) {\( s\)};
    \vertex[below=2em of c3] (c1) {\(\bar s\)};
    \vertex at ($(c1)!0.5!(c3) - (1cm, 0)$) (c2);
    
    \vertex[above=4em of a3] (d1) {\( u\)};
    \vertex[below=2em of d1] (d3) {\(\bar d\)};
    \vertex at ($(d1)!0.5!(d3) - (1cm, 0)$) (d2);
    
    \vertex[below=5em of b1] (e1) {\(d\)};
    \vertex[right=2cm of e1] (e2);
    \vertex[right=2cm of e2] (e3) {\(d\)};
    
    \diagram* {
      {[edges=fermion]
        (a1) -- (a2) -- (a3),
        (b1) -- (b2) -- (b3),
        (e1) -- (e2) -- (e3),
      },
      (a2) -- [boson,bend left, edge label=\(W^+\)] (d2),

      (c1) -- [fermion, out=180, in=-45] (c2) -- [fermion, out=45, in=180] (c3),
      (d3) -- [fermion, out=180, in=-45] (d2) -- [fermion, out=45, in=180] (d1),
    };

    \draw [decoration={brace}, decorate] (e1.south west) -- (a1.north west)
          node [pos=0.5, left] {\(\Lambda_{c}^{+}\)};
    \draw [decoration={brace}, decorate] (a3.north east) -- (c3.south east)
          node [pos=0.5, right] {\(\Sigma^{0}\)};
    \draw [decoration={brace}, decorate] (d1.north east) -- (d3.south east)
         node [pos=0.5, right] {\(\pip\)};
    \draw [decoration={brace}, decorate] (c1.north east) -- (e3.south east)
         node [pos=0.5, right] {\(K^{0}\)};
  \end{feynman}
  \end{tikzpicture}
\end{adjustbox}
\put(-80,-15){(c)}
\end{minipage}
\begin{minipage}[b]{.32\linewidth}
  \begin{adjustbox}{width=\linewidth}
\begin{tikzpicture}
  \begin{feynman}

    \vertex (a1) {\( u\)};
    \vertex[right=2cm of a1] (a2);
    \vertex[right=2cm of a2] (a3){\(u\)};

    \vertex[below=4em of a1] (b1) {\(d\)};
    \vertex[right=2cm of b1] (b2);
    \vertex[right=2cm of b2] (b3) {\(u\)};

    \vertex[below=1em of a3] (c3) {\(\bar s\)};
    \vertex[below=2em of c3] (c1) {\( s\)};
    \vertex at ($(c1)!0.5!(c3) - (1cm, 0)$) (c2);
    
    \vertex[below=1em of b3] (d1) {\( d\)};
    \vertex[below=2em of d1] (d3) {\(\bar d\)};
    \vertex at ($(d1)!0.5!(d3) - (1cm, 0)$) (d2);
    
    \vertex[below=5em of b1] (e1) {\(c\)};
    \vertex[right=2cm of e1] (e2);
    \vertex[right=2cm of e2] (e3) {\(s\)};
    
    \diagram* {
      {[edges=fermion]
        (a1) -- (a2) -- (a3),
        (b1) -- (b2) -- (b3),
        (e1) -- (e2) -- (e3),
      },
    (e2) -- [boson,right, edge label=\(W^+\)] (b2),

      (c3) -- [fermion, out=-180, in=45] (c2) -- [fermion, out=-45, in=180] (c1),
      (d3) -- [fermion, out  =180, in=-45] (d2) -- [fermion, out=45, in=180] (d1),
    };

    \draw [decoration={brace}, decorate] (e1.south west) -- (a1.north west)
          node [pos=0.5, left] {\(\Lambda_{c}^{+}\)};
    \draw [decoration={brace}, decorate] (c1.north east) -- (d1.south east)
          node [pos=0.5, right] {\(\Sigma^{0}\)};
    \draw [decoration={brace}, decorate] (a3.north east) -- (c3.south east)
         node [pos=0.5, right] {\(K^{+}\)};
    \draw [decoration={brace}, decorate] (d3.north east) -- (e3.south east)
         node [pos=0.5, right] {\(\bar K^{0}\)};
  \end{feynman}
  \end{tikzpicture}
    \end{adjustbox}
   \put(-80,-15){(d)}
\end{minipage}
\begin{minipage}[b]{.32\linewidth}
  \begin{adjustbox}{width=\linewidth}
\begin{tikzpicture}

  \begin{feynman}

    \vertex (a1) {\( u\)};
    \vertex[right=2cm of a1] (a2);
    \vertex[right=2cm of a2] (a3){\(u\)};

    \vertex[below=4em of a1] (b1) {\(d\)};
    \vertex[right=2cm of b1] (b2);
    \vertex[right=2cm of b2] (b3) {\(d\)};

    \vertex[below=1em of a3] (c3) {\(\bar s\)};
    \vertex[below=2em of c3] (c1) {\( s\)};
    \vertex at ($(c1)!0.5!(c3) - (1cm, 0)$) (c2);
    
    \vertex[below=1em of b3] (d1) {\( u\)};
    \vertex[below=2em of d1] (d3) {\(\bar d\)};
    \vertex at ($(d1)!0.5!(d3) - (1cm, 0)$) (d2);
    
    \vertex[below=5em of b1] (e1) {\(c\)};
    \vertex[right=2cm of e1] (e2);
    \vertex[right=2cm of e2] (e3) {\(s\)};
    
    \diagram* {
      {[edges=fermion]
        (a1) -- (a2) -- (a3),
        (b1) -- (b2) -- (b3),
        (e1) -- (e2) -- (e3),
      },
      (e2) -- [boson,bend left, edge label=\(W^+\)] (d2),

      (c3) -- [fermion, out=-180, in=45] (c2) -- [fermion, out=-45, in=180] (c1),
      (d3) -- [fermion, out=180, in=-45] (d2) -- [fermion, out=45, in=180] (d1),
    };

    \draw [decoration={brace}, decorate] (e1.south west) -- (a1.north west)
          node [pos=0.5, left] {\(\Lambda_{c}^{+}\)};
    \draw [decoration={brace}, decorate] (c1.north east) -- (d1.south east)
          node [pos=0.5, right] {\(\Sigma^{0}\)};
    \draw [decoration={brace}, decorate] (a3.north east) -- (c3.south east)
         node [pos=0.5, right] {\(K^{+}\)};
    \draw [decoration={brace}, decorate] (d3.north east) -- (e3.south east)
         node [pos=0.5, right] {\(\bar K^{0}\)};
  \end{feynman}
  \end{tikzpicture}
\end{adjustbox}
\put(-80,-15){(e)}
\end{minipage}
\caption{The (a) W-exchange diagram, (b) internal $W$-emission diagram, (c) external $W$-emission diagram for $\Lambda_c^+ \to \Sigma^0 K_S^0 \pi^+$ as well as (d) $W$-exchange diagram and (e) internal $W$-emission diagram for $\Lambda_c^+ \to \Sigma^0 K_S^0 K^+$.}\label{fig:feynmand}
  \end{figure}

The total BF of the measured $\Lambda_c^+$ decays is still only around
70\% according to the Particle Data Group
(PDG)~\cite{ParticleDataGroup:2024cfk}, and many decay modes of
$\Lambda_c^+$ remain unknown. An upper limit for the BF of $\Lambda_c^+ \to \Sigma^0
K_S^0 K^+$ has been determined by BESIII using the 4.5 fb$^{-1}$ of data
collected at center-of-mass energies between 4.600 and 4.700 GeV with
the double tag method to be $1.28\times10^{-3}$ at the
90\% confidence level (C.L.)~\cite{BESIII:2025rda}. In this work, we use more data samples with single tag method to investigate this decay mode again.  The decay $\Lambda_c^+ \to \Sigma^0 K_S^0 \pi^+$ has not been observed
experimentally. Both decays involve the $W$-exchange and $W$-emission
diagrams as shown in Fig.~\ref{fig:feynmand}.

The $\Lambda_c$ decays can have contributions from both direct
nonresonant and resonant processes. The nonresonant contributions to
$\Lambda_c^+ \to \Sigma^0 K_S^0 \pi^+$ and $\Lambda_c^+ \to \Sigma^0
K_S^0 K^+$ have been investigated using
$SU(3)_F$~\cite{Geng:2018upx,Geng:2024sgq}, with the predicted BFs
listed in Table~\ref{tab:theo}. In addition, the decay $\Lambda_c^+
\to \Sigma^0 K_S^0 K^+$ can proceed via the intermediate channel
$\Lambda_c^+\to \Sigma^0 a_0(980)^+$~\cite{Wang:2025uie}, followed by
$a_0(980)^+\to K^0_S K^+$.  This opens a new avenue for probing light
scalar mesons, especially taking into account the BESIII measurement
of
 ${\cal B}(\Lambda_c^+\to \Lambda a_0(980)^+)$~\cite{BESIII:2024mbf},
 which significantly exceeds theoretical predictions.  Similarly, the
 decay $\Lambda_c^+ \to \Sigma^0 K_S^0 \pi^+$ may involve various
 resonant decays, such as $\Lambda_c^+ \to \Sigma^0 K^{*+}(K^{*+} \to K_S^0 \pi^+), K_S^0 \Sigma^{*0} (\Sigma^{*0} \to \Sigma^0 \pi^+)$, and $\Sigma^0K_0^*(700)^+(K_0^*(700)^+\to K^0_S \pi^+)$.  Differences from
 theoretical predictions in these decays can provide valuable insights
 into the presence of additional resonant contributions. For instance,
 BESIII has observed, in addition to a nonresonant component, the decay
 $\Lambda_c^+ \to \Lambda K^*(892)^+$~\cite{BESIII:2024xny} through an analysis of $\Lambda_c^+ \to
 \Lambda K_S^0 \pi^+$. Moreover, the reported ${\cal B}(\Lambda_c^+
 \to \Lambda K^*(892)^+)$ is consistent with the theoretical
 predictions based on $SU(3)_F$~\cite{Hsiao:2019yur,Geng:2020zgr}.

In this work, we measure the BFs of $\Lambda_c^+ \to \Sigma^0 K_S^0
\pi^+$ and $\Lambda_c^+ \to \Sigma^0 K_S^0 K^+$ by analyzing
6.4~fb$^{-1}$ of data taken at 13 energy points between $\sqrt{s}=4.600$
and $4.950$~GeV~\cite{BESIII:2022ulv} with the BESIII detector at the
BEPCII collider. The charge-conjugate processes are implicitly
included throughout this paper.
  
\begin{table}[h!]
\centering
\caption{\label{tab:theo} Theoretical results for nonresonant decays
based on $SU(3)_F$~\cite{Geng:2024sgq}.}
\begin{tabular}{l|c|c} \hline \hline
Decay mode &$\Lambda_c^+\to\Sigma^0K_S^{0}\pip$ &
$\Lambda_c^+\to\Sigma^{0}K_S^{0}K^{+}$ \\ \hline BF
&$(0.17\pm0.05)\times10^{-3}$ &$(0.12\pm0.04)\times10^{-3}$ \\ \hline
\hline
\end{tabular}
\end{table}

\section{BESIII DETECTOR AND MONTE CARLO SIMULATION}
\label{sec:detector}
\hspace{1.5em}

The BESIII detector~\cite{Ablikim:2009aa} records symmetric $e^+e^-$
collisions provided by the BEPCII storage
ring~\cite{Yu:IPAC2016-TUYA01}, which operates with a peak luminosity
of $1.1\times10^{33}$~cm$^{-2}$s$^{-1}$ in the center-of-mass energy
range from 1.84 to 4.95~GeV.  BESIII has collected large data samples
in this energy region~\cite{Ablikim:2019hff}. The cylindrical core of
the BESIII detector covers 93\% of the full solid angle and consists
of a helium-based
 multilayer drift chamber~(MDC), a plastic scintillator time-of-flight
system~(TOF), and a CsI(Tl) electromagnetic calorimeter~(EMC),
which are all enclosed in a superconducting solenoidal magnet
providing a 1.0~T magnetic field. The solenoid is supported by an
octagonal flux-return yoke with resistive plate counter muon
identification modules interleaved with steel. 

The charged-particle momentum resolution at $1~{\rm GeV}/c$ is
$0.5\%$, and the $dE/dx$ resolution is $6\%$ for electrons from Bhabha
scattering. The EMC measures photon energies with a resolution of
$2.5\%$ ($5\%$) at $1$~GeV in the barrel (end cap) region. The time
resolution in the TOF barrel region was 68~ps, while that in the end
cap region is 110~ps. The end cap TOF system was upgraded in 2015
using multi-gap resistive plate chamber technology, providing a time
resolution of 60~ps~\cite{etof}. About 90\% of the data used were
collected after this upgrade.

Simulated data samples produced with the {\sc
geant4}-based~\cite{geant4} MC software, which includes the geometric
and material description of the BESIII detector and the detector
response, are used to determine detection efficiencies and to estimate
backgrounds. The simulation models considered the beam energy spread
and initial state radiation (ISR) in the $e^+e^-$ annihilations with
the generator {\sc kkmc}~\cite{ref:kkmc}. The inclusive MC sample
includes the production of open charm processes, the ISR production of
vector charmonium(-like) states, and the continuum processes
incorporated in {\sc kkmc}.

The known decay modes are modeled with {\sc evtgen}~\cite{ref:evtgen}
using BFs taken from the PDG~\cite{ParticleDataGroup:2024cfk}, and the
remaining unknown charmonium decays are modeled with {\sc
lundcharm}~\cite{ref:lundcharm}. Final state radiation from charged
final state particles is incorporated using {\sc photos}~\cite{photos2}.

For the MC production of the $e^+e^-\rightarrow
\Lambda_c^+\bar{\Lambda}_c^-$ events, the observed cross sections are
taken into account, and the signal decay processes are modeled to be
uniformly distributed in phase space (PHSP). All the final tracks and
photons are reconstructed by the {\sc geant4}-based~\cite{geant4}
detector simulation package.

\section{EVENT SELECTION AND DATA ANALYSIS}
\label{sec:analysis}
\hspace{1.5em} 

To reconstruct the signal processes, all the final particles are
selected to form the $\Lambda_c^+$ candidates.  Charged tracks
detected in the MDC are required to be within a polar angle ($\theta$)
range of $|\rm{cos\theta}|<0.93$, where $\theta$ is defined with
respect to the $z$-axis, which is the symmetry axis of the MDC. For
charged tracks not originating from $K_S^0$ or $\Lambda$ decays, the
distance of closest approach to the interaction point (IP) must be
less than 10\,cm along the $z$-axis, $|V_{z}|$, and less than 1\,cm in
the transverse plane, $|V_{xy}|$.

Photon candidates are identified using isolated showers in the EMC.
The deposited energy of each shower must be more than 25~MeV in the
barrel region ($|\cos \theta|< 0.80$) and more than 50~MeV in the end
cap region ($0.86 <|\cos \theta|< 0.92$).  To exclude showers that
originate from charged tracks, the angle subtended by the EMC shower
and the position of the closest charged track at the EMC must be
greater than 10 degrees as measured from the IP. To suppress
electronic noise and showers unrelated to the event, the difference
between the EMC time and the event start time is required to be within
[0, 700]\,ns.
  
Particle identification~(PID) for charged tracks combines measurements
of the energy deposited in the MDC~(d$E$/d$x$) and the flight time in
the TOF to form likelihoods $\mathcal{L}(h)~(h=p,K,\pi)$ for each
hadron $h$ hypothesis.  Tracks are identified as protons when the
proton hypothesis has the greatest likelihood
($\mathcal{L}(p)>\mathcal{L}(K)$ and
$\mathcal{L}(p)>\mathcal{L}(\pi)$), then charged kaons and pions are
identified by comparing the likelihoods for the kaon and pion
hypotheses, $\mathcal{L}(K)>\mathcal{L}(\pi)$ and
$\mathcal{L}(\pi)>\mathcal{L}(K)$, respectively.

The $\Sigma^0$ is reconstructed through its decay final state
$\gamma\Lambda$($\Lambda\to p\pi^-$). The invariant mass of the
$M(\gamma\Lambda)$ is required to be within (1.179,~1.203)
$\mathrm{GeV}/c^{2}$.

Each $\Lambda$($K_{S}^0$) candidate is reconstructed from two
oppositely charged tracks satisfying $|V_{z}|<$ 20~cm and
$|\rm{cos\theta}|<0.93$.  The two charged tracks are assigned as
$p\pi^{-}$($\pi^+\pi^-$) and only the proton is subjected to further
PID criteria. They are constrained to originate from a common vertex
by applying a vertex fit and requiring the $\chi^2$ of a vertex fit to
be less than 100. The invariant mass is required to satisfy the
$1.111~<M(p\pi^{-})< 1.121$~GeV/$c^{2}$($0.487~<M(\pi^{+}\pi^{-})<
0.511$~GeV/$c^{2}$), which corresponds to three times the standard
deviation of the reconstruction resolution around the known
$\Lambda$($K_S^0$) mass~\cite{ParticleDataGroup:2024cfk}. The decay
length of the $\Lambda$($K_{S}^0$) candidate is required to be greater
than twice the vertex resolution away from the IP.

The signal candidates are identified using the beam constrained mass
$M_{\rm BC} = \sqrt{E_{\rm beam}^2/c^4 - p^2/c^2}$, where $p$ is the
momentum of reconstructed $\Lambda_c^+$ candidates in the
center-of-mass frame of the $e^+e^-$ collision. To improve the signal
purity, a four-constraint (4C) kinematic fit is performed for
$\Lambda_c^+$ candidates. Here, we constrain the invariant mass of
recoil side of $\Lambda_c^+$ candidates to the $\bar{\Lambda}_c^-$
nominal mass~\cite{ParticleDataGroup:2024cfk}, and the invariant masses of
$\pi^+\pi^-$, $p\pi^-$ and $\gamma p\pi^-$ to the
nominal masses of $K_S^0$, $\Lambda$ and $\Sigma^0$. The
candidate with the minimum $\chi^2_{4\rm C}$ given by the $4 \rm C$
kinematic fit is retained for further analysis. The $\chi^2_{4\rm
C}$ requirements are optimized by using the Figure of Merit
(FOM$=S/\sqrt{S+B})$, where $S$ is the number of signal events in the region
$M_{\rm BC} \in [2.282,~2.291]$ GeV/$c^2$ generated by the theoretical
predicted BF and updated by our final result, $B$
is the number of background events estimated using the inclusive MC, and
$S$ and $B$ are normalized to the integrated luminosity of the
data sample. We require $\chi^2_{4\rm C} < 29$ for
$\Lambda_c^+\to\Sigma^0K_{S}^{0}\pip$ and $\chi^2_{4\rm C} < 171$ for
$\Lambda_c^+\to\Sigma^0K_{S}^{0}K^{+}$.

\begin{figure*}[h!]
  \centering

\includegraphics[width=0.34\paperwidth]{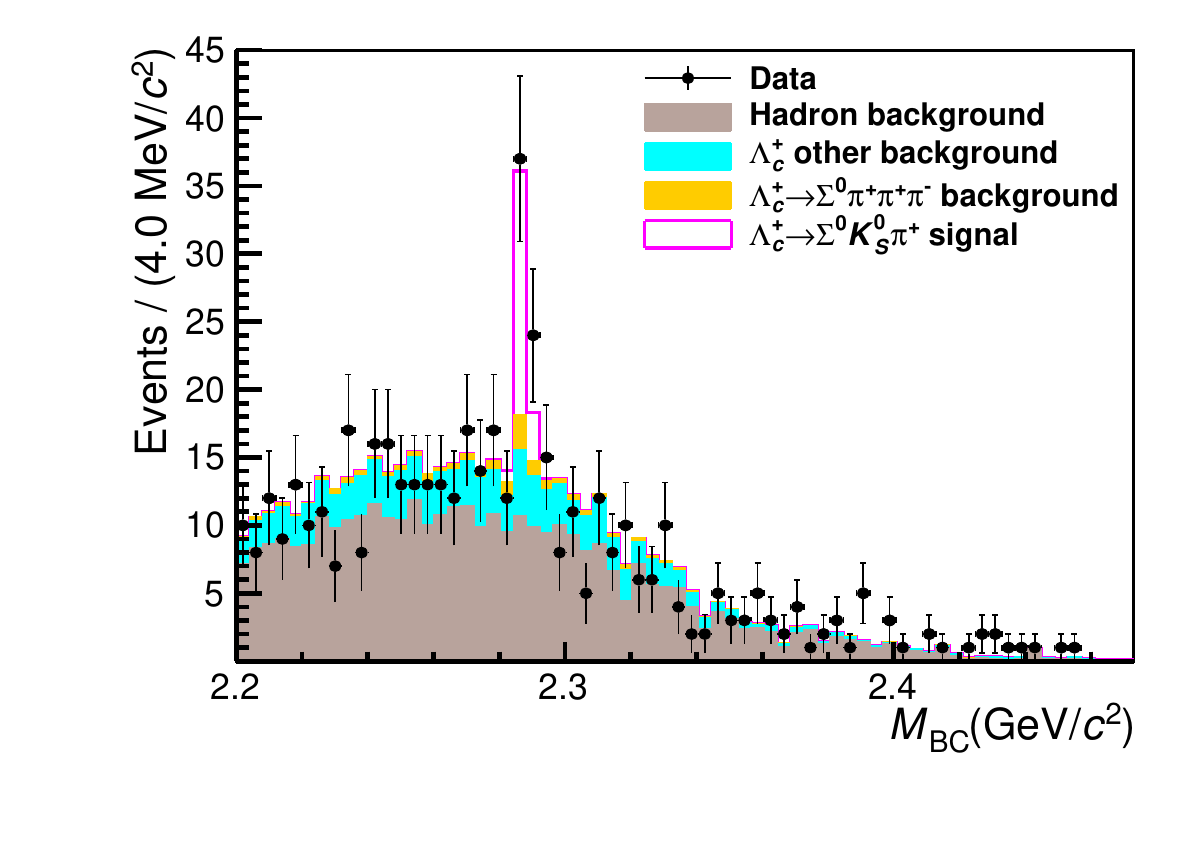}
 \put(-160,120){(a)}
\includegraphics[width=0.34\paperwidth]{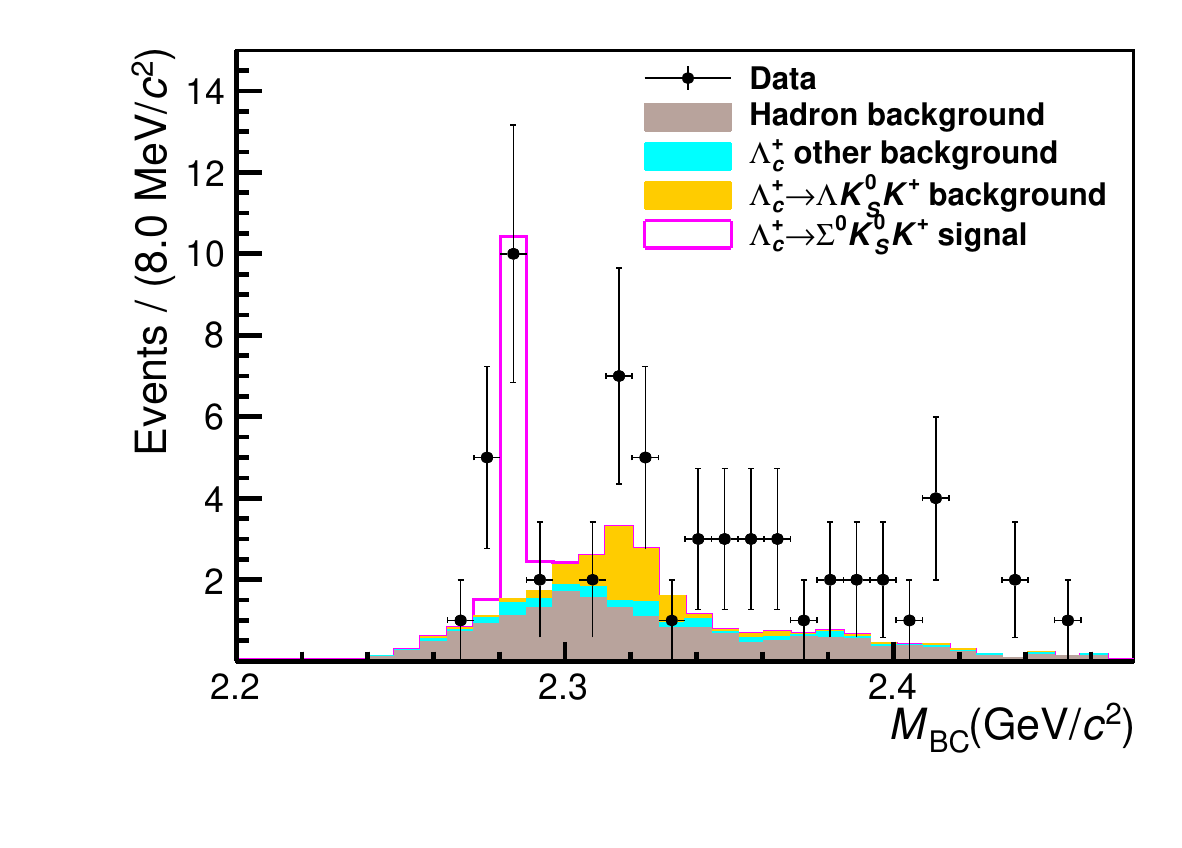}
\put(-160,120){(b)}
\caption {The $M_{\rm BC}$ distributions of the accepted candidates
for (a) $\Lambda_c^+\to\Sigma^0K_{S}^{0}\pip$ and (b)
$\Lambda_c^+\to\Sigma^0K_{S}^{0}K^{+}$. The black points with error
bars are data, and the histograms are MC simulated events. The
 magenta, yellow, cyan, and brown histograms
 are the signal, peaking
backgrounds, $\Lambda_c^+$ backgrounds and
combinatorial backgrounds, respectively.}
\label{fig:BKG}

\end{figure*}

After all the selection criteria, the $M_{\rm BC}$ distributions for
both data and inclusive MC samples are shown in
Fig.~\ref{fig:BKG}. Some peaking backgrounds
($\Lambda_c^+\to\Sigma^0\pip\pi^-\pip$ in the
$\Lambda_c^+\to\Sigma^0K_{S}^{0}\pip$ analysis and
$\Lambda_c^+\to\Lambda K_{S}^{0}K^+$ in the
$\Lambda_c^+\to\Sigma^0K_{S}^{0}K^+$ analysis) are observed.

\section{DETERMINATION OF THE BRANCHING FRACTION}
\label{sec:result}
\hspace{1.5em}
At each energy point, the BF of the signal decay is calculated by

\begin{align}
 \mathcal{B}&\equiv\frac{N^{\rm sig}}{2\cdot N_{\Lambda_c^+\bar{\Lambda}_c^-}\cdot\varepsilon^{\rm sig}\cdot\mathcal{B}^{\rm inter}},
 \tag{1} \label{eq:BF2} \end{align}
where $N^{\rm sig}$ denotes the signal yield, 2 is because the charge
conjugate process also contributes, $N_{\Lambda_c^+\bar{\Lambda}_c^-}$
is the
 number of $\Lambda_c^+\bar{\Lambda}_c^-$ events listed in Table~\ref{tab:efficiency} and  calculated using the
 cross section~\cite{BESIII:2023rwv} and integrated
 luminosity~\cite{BESIII:2015qfd,BESIII:2022ulv}, $\varepsilon^{\rm
 sig}$ is the signal efficiency from the MC simulation shown in
 Table~\ref{tab:efficiency} and $\mathcal{B}^{\rm inter}$ is
 determined by the BFs of the intermediate states
 ($\Sigma^0\to\gamma\Lambda$ $(100\%)$, $\Lambda\to p\pi^-$ $(63.90\pm0.05)\%$ and
 $K_S^0\to\pi^+\pi^-$ $(69.20\pm 0.05)\%$))~\cite{ParticleDataGroup:2024cfk}.

\begin{table}[h!]
\centering
\caption{\label{tab:efficiency} The detection efficiencies for $\Lambda_c^+\to\Sigma^0K_{S}^{0}\pip$ and $\Lambda_c^+\to\Sigma^0K_{S}^{0}K^{+}$, where the uncertainties are statistical only, and the number of $\Lambda^+_c\bar{\Lambda}^-_c$ events. The efficiencies do not include the BF of the sequential decay of $K_{S}^{0}$. }
\begin{tabular}{cccc } 
\hline
\hline
$\sqrt{s}(\mev)$  &$\varepsilon_{\Lambda_c^+\to\Sigma^0K_{S}^{0}\pip}(\%)$&$\varepsilon_{\Lambda_c^+\to\Sigma^0K_{S}^{0}K^{+}}(\%)$ & $N_{\Lambda^+_c\bar{\Lambda}^-_c}$\\ 
\hline
4599.53 &6.52$\pm$0.04 &1.97$\pm$0.02 & 99244\\
\hline                                                                                            
4611.86 &5.78$\pm$0.04 &1.87$\pm$0.02 & 17442\\
\hline                                                                                            
4628.00 &5.58$\pm$0.03 &2.03$\pm$0.02 & 89280\\
\hline                                                                                           
4640.91 &5.59$\pm$0.04 &2.27$\pm$0.02 & 95426\\
\hline                                                                                            
4661.24 &5.59$\pm$0.04 &2.46$\pm$0.02 & 91639\\
\hline                                                                                           
4681.92 &5.60$\pm$0.04 &2.66$\pm$0.02 & 278622 \\
\hline                                                                                            
4698.82 &5.45$\pm$0.04 &2.75$\pm$0.02 & 84341\\
\hline
4739.70 &5.96$\pm$0.03 &3.28$\pm$0.03 & 19848\\
\hline                                                                                            
4750.05 &5.76$\pm$0.04 &3.38$\pm$0.03 &  45091\\
\hline                                                                                           
4780.54 &5.70$\pm$0.04 &3.57$\pm$0.03 & 61431 \\
\hline                                                                                            
4843.07 &5.14$\pm$0.03 &3.66$\pm$0.03 & 45429\\
\hline                                                                                           
4918.02 &4.46$\pm$0.03 &3.58$\pm$0.03 & 20634\\
\hline                                                                                            
4950.93 &4.06$\pm$0.03 &3.32$\pm$0.03 & 14416\\
\hline
\hline
\end{tabular}
\end{table}

\begin{figure*}[h!]
  \centering \includegraphics[width=0.34\paperwidth]{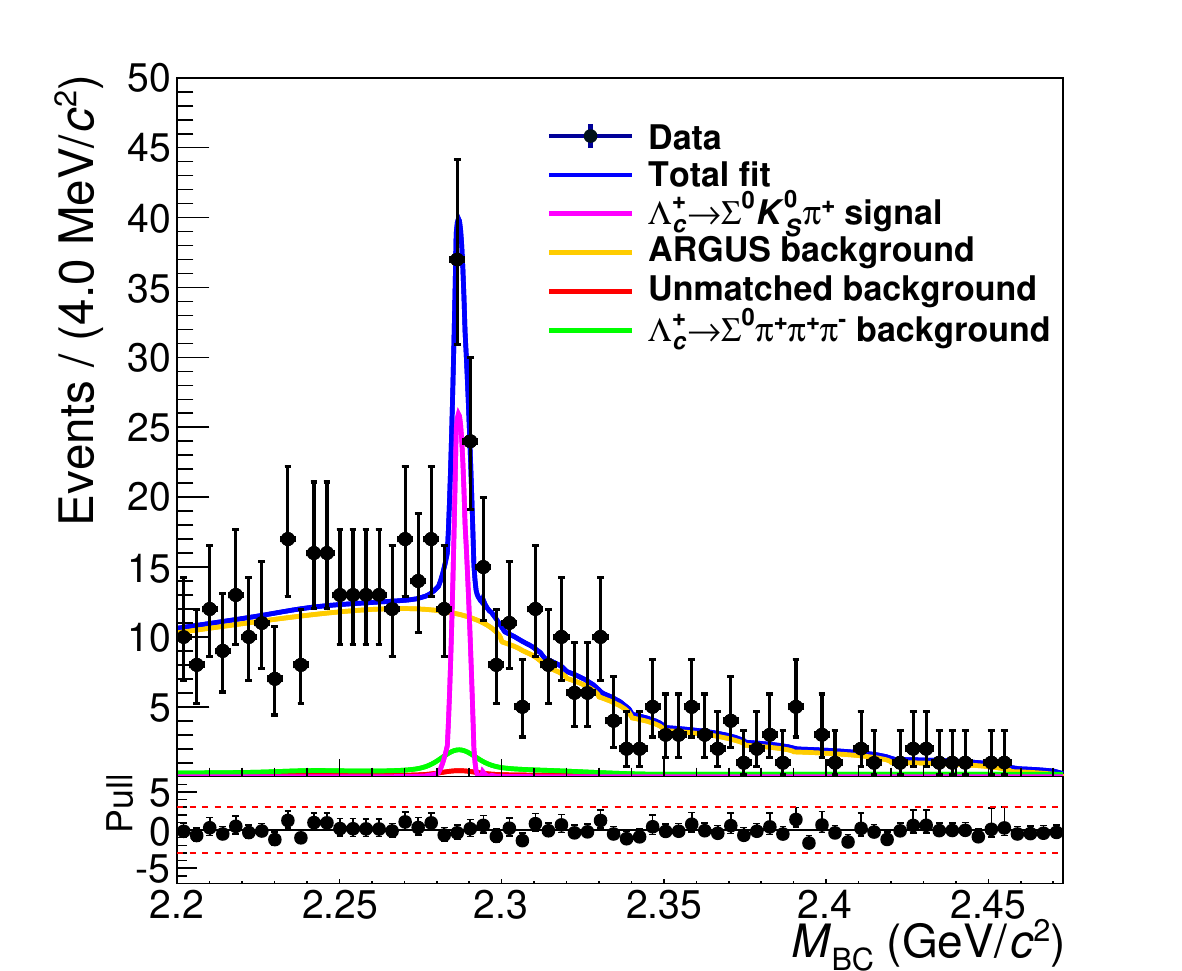}
  \put(-160,140){(a)}
  \includegraphics[width=0.34\paperwidth]{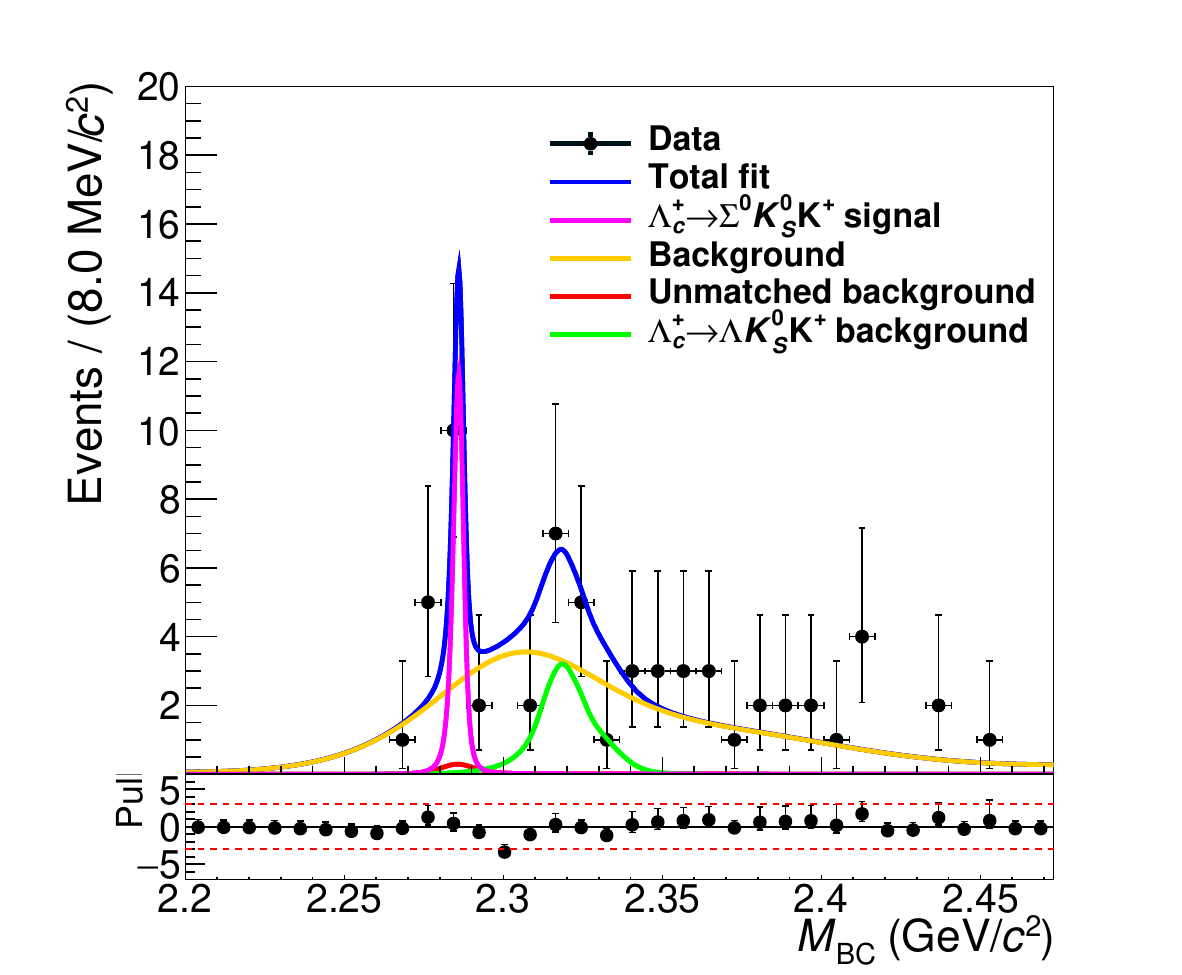}
  \put(-160,140){(b)}
\caption {Results of the fits to the $M_{BC}$
  distributions for (a) $\Lambda_c^+\to\Sigma^0K_{S}^{0}\pip$ and (b)
  $\Lambda_c^+\to\Sigma^0K_{S}^{0}K^{+}$ combining 13 energy
  points. The black points with error bars are data,
 the blue, yellow, magenta, green, and red solid lines 
 are the
total fit, the ARGUS background
function ($\Lambda_c^+ \to \Sigma^0 K_S^0 \pi^+$) and the MC
background shape ($\Lambda_c^+ \to \Sigma^0 K_S^0 K^+$), the signals, the peaking
backgrounds, and the unmatched components, respectively.}
\label{fig:fit}
\end{figure*}

For $\Lambda_c^+\to\Sigma^0K_{S}^{0}\pip$, the maximum likelihood fit
is performed simultaneously on the $M_{\rm BC}$ distributions of all
the energy points. In the fit, the BF value is constrained to be the
same at each energy point. The MC histograms for the signal shapes in
Fig.\ref{fig:BKG} also contains some mis-reconstructed photons and
mis-reconstructed $\pi^+$s and $K^+$s. To obtain clean signal shapes,
a match based on the MC truth information is performed for signal
processes using the inclusive MC sample.  The angle between the
reconstructed and the truth three-momenta of the tracks is required to
be less than 10 degrees. The others are classified as unmatched
background and provide the unmatched shape.  The clean signal shapes
are convolved with Gaussian functions, which account for the mass
resolution difference between data and MC simulation. The parameters
of the Gaussian functions are floated, but are constrained to
be the same at each energy point. The background shapes are described
with ARGUS functions~\cite{ARGUS:1990hfq} with floating parameters,
except for the endpoints which are fixed by the center-of-mass
energy. The ratio of unmatched backgrounds with respect to the signal
yields is fixed. For the peaking backgrounds
$\Lambda_c^+\to\Sigma^0\pip\pi^-\pip$, the yields are fixed using
BFs from the PDG~\cite{ParticleDataGroup:2024cfk}, and the shapes and
efficiencies are obtained from the MC sample. By considering
$N_{\Lambda_c^+\bar{\Lambda}_c^-}$, the efficiencies and BFs of
intermediate states, the BF of $\Lambda_c^+\to\Sigma^0K_{S}^{0}\pip$
can be obtained.  The sums of the fit plots of all energy points are
shown in Fig.~\ref{fig:fit}.

\begin{figure*}[h!]
  \centering \includegraphics[width=0.34\paperwidth]{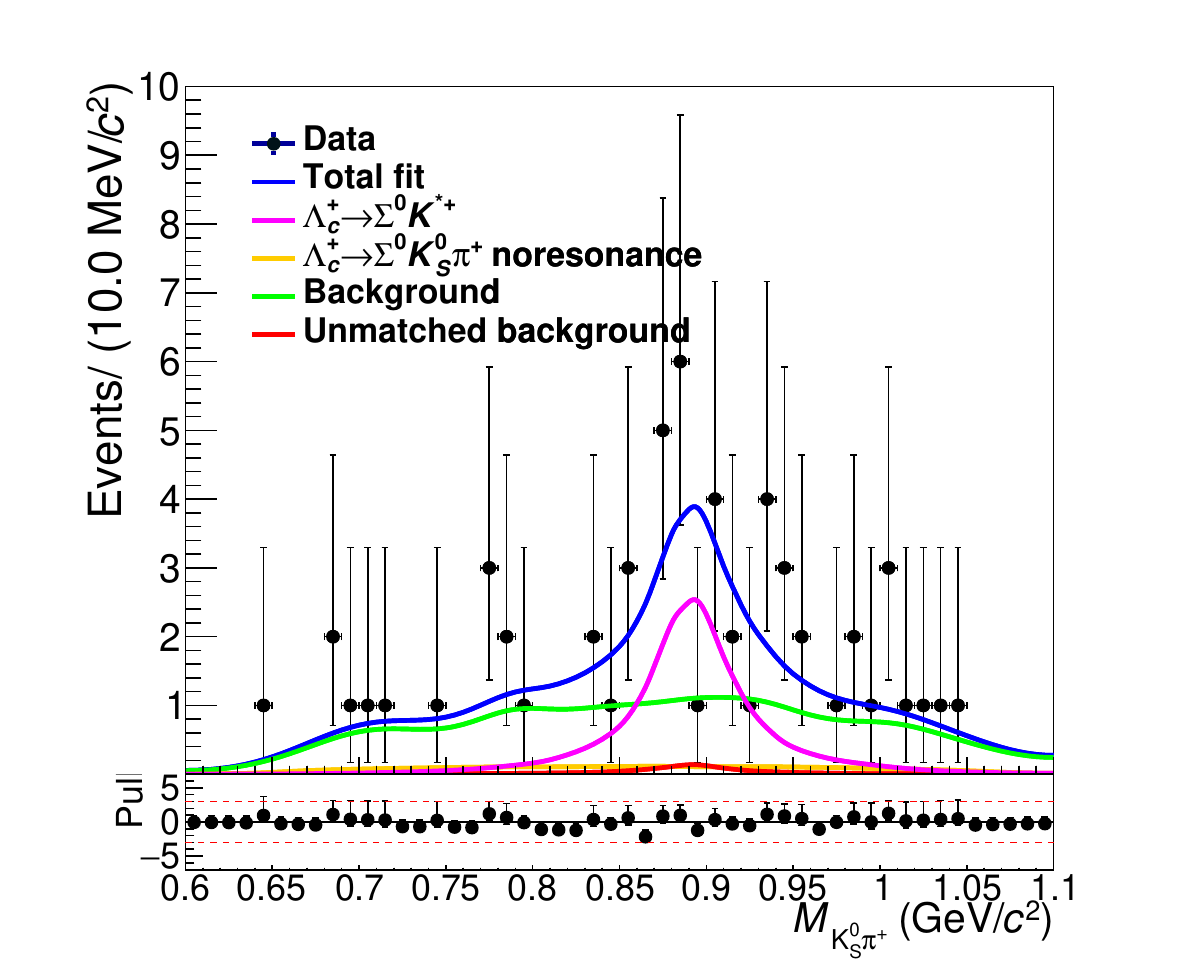}
  \caption {The fit plot of $\Lambda_c^+\to\Sigma^0 K^{*+}$ combining
  13 energy points in the $M_{\rm BC}$ signal region (2.282, 2.291)
  GeV/$c^2$. The black points with error bars are data, and the blue,
  magenta, yellow, red, and green lines are the total fit, signal,
  $\Lambda _c^+ \to \Sigma^0 K_S^0 \pi^+$(non-$K^{*+}$) process,
  unmatched component, and other background, respectively.}
  \label{fig:KSpi}
\end{figure*}

The $K_S^0\pi^+$‌invariant mass distribution combining 13
energy points is shown in Fig.~\ref{fig:KSpi}. The
$\Lambda_c^+\to\Sigma^0 K^{*+}$ decay is seen, and the BF result is
obtained by fitting the $M_{K_S^0\pi^+}$ distribution in the $M_{\rm
BC}$ signal region (2.282, 2.291) GeV/$c^2$. Included in the fit are
the contributions of $\Lambda_c^+\to\Sigma^0 K^{*+}$,
$\Lambda_c^+\to\Sigma^0K_{S}^{0}\pip$(non-$K^{*+}$), unmatched
backgrounds and other backgrounds, and the shapes are obtained from MC
samples. The signal shape is convolved with a Gaussian function. The
mean and sigma values of the Gaussian are floated. The yields of
unmatched backgrounds and other backgrounds are fixed using the result
of the $M_{\rm BC}$ fit, and the yields of other components are
floated. The efficiency is ($5.31\pm0.03$)\%.


For $\Lambda_c^+\to\Sigma^0K_{S}^{0}K^{+}$, due to the ‌limited
statistics, all the data samples are combined in the fit process.  The
clean signal shape and unmatched background shape are obtained as for
the $\Lambda_c^+\to\Sigma^0K_{S}^{0}\pip$ case. The parameters of the
Gaussian function convolved with the clean signal shape are fixed to
the values of $\Lambda_c^+\to\Sigma^0K_{S}^{0}\pip$. The background
shape is described by the inclusive MC sample. The ratio of the
unmatched background to the signal yield is fixed. For the peaking
background $\Lambda_c^+\to\Lambda K_S^0 K^+$, the yield is fixed by
using the PDG BF~\cite{ParticleDataGroup:2024cfk}, while the efficiency and
shape are obtained from the MC sample.

\begin{figure*}[h!]
\centering
	\includegraphics[width=0.34\paperwidth]{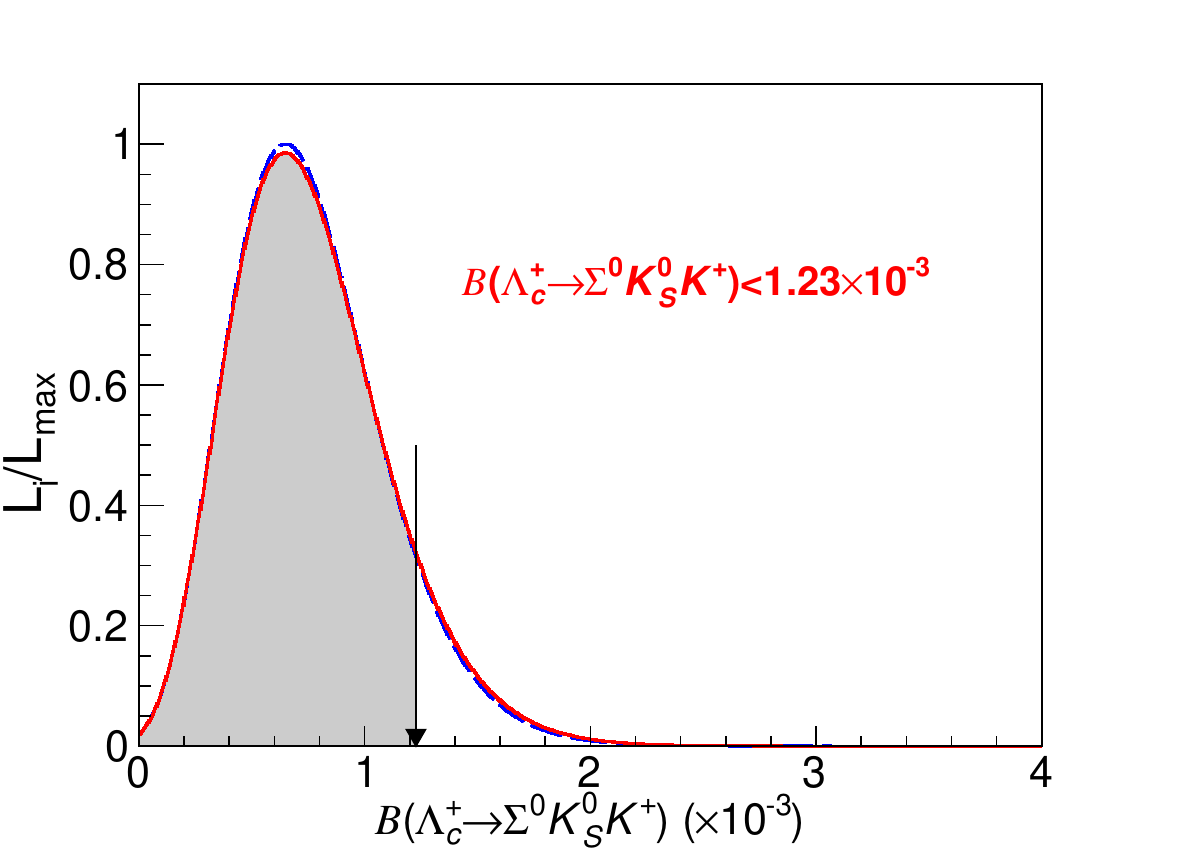}
\caption{The likelihoods curve of $\Lambda_c^+\to\Sigma^0K^0_SK^+$. The results obtained with and without incorporating the systematic uncertainties are shown in the red solid and blue dashed curves, respectively. The black arrow shows the result corresponding to the 90\% C.L. }
\label{fig:prob}
\end{figure*}

The signal yields, the BF results and the significances are shown in
Table~\ref{tab:BF_yields}. Since the statistical significance of
$\Lambda_c^+\to\Sigma^0K_{S}^{0} K^+$ is only 3.7$\sigma$ by comparing
the fit likelihoods with and without including the signal components,
the upper limit is determined by a likelihood scan which takes into
account the systematic uncertainties.  The likelihood curve calculated
is shown in Fig.~\ref{fig:prob}. The upper limit at the 90\% C.L. is
$(1.23\times10^{-3})$.

\begin{table}[h!] 
\centering
\caption{\label{tab:BF_yields} The signal yield, BF and significance for each decay mode, where the uncertainties are statistical only. For $\Lambda_c^+\to\Sigma^0K_{S}^{0}\pi^{+}$ and $\Lambda_c^+\to\Sigma^0K_{S}^{0}K^{+}$, the BF contains all resonance. For $\Lambda_c^+\to\Sigma^0K^{*+} \to \Sigma^0K_{S}^{0}\pi^{+}$, BF is $\mathcal{B}(\Lambda_c^+
\to \Sigma^0K^{*+}) \times \mathcal{B}(K^{*+} \to K^0_S \pi^+)$.}
\begin{tabular}{lcccc}
\hline
\hline
Decay mode  &$N_{\rm sig}$&BFs($\times 10^{-3}$) & Statistical significance  \\
\hline
$\Lambda_c^+\to\Sigma^0K_{S}^{0}\pip$&$28.24\pm7.15$ & $0.58\pm 0.14$  & $5.9\sigma$ \\
\hline              
$\Lambda_c^+\to\Sigma^0K^{*+} \to \Sigma^0K_{S}^{0}\pi^{+}$ & $18.48\pm8.51$ & $0.41\pm0.19$ & $2.9\sigma$ \\
\hline
$\Lambda_c^+\to\Sigma^0K_{S}^{0}K^{+}$&$7.92\pm3.43$    & $0.35\pm 0.16$  & $3.7\sigma$\\
\hline
\hline
\end{tabular}
\end{table}

%

\newpage
\section{SYSTEMATIC UNCERTAINTY}
\label{sec:systematic}
\hspace{1.5em}

The systematic uncertainties on the BF measurements include
contributions from the tracking, PID, $K_S^0$ and $\Lambda$
reconstruction, photon selection, $\mathcal{B}^{\rm inter}$,
$\chi^2_{4C}$ kinematic fit requirement, the fit process, MC model,
the peaking and unmatched background estimation, and the number of
$N_{\Lambda_c^+\bar\Lambda_c^-}$ events. They are summarized in
Table~\ref{tab:systemUncertain} and the details are described
below. The total systematic uncertainties are calculated as the sum in
quadrature of the individual contributions by assuming the sources are
independent of one another.

\begin{table}[h!]
\centering\small
\caption{\label{tab:systemUncertain} The systematic uncertainties ($\%$). }
\begin{tabular}{l|c|c}
\hline
\hline
Source&$\Lambda_c^+\to\Sigma^0K_{S}^{0}\pip$&$\Lambda_c^+\to\Sigma^0K_{S}^{0}K^{+}$ \\ \hline
 Tracking & 1.0 & 2.0               \\
 PID & 1.0 & 2.0               \\
$K_{S}^{0}$ reconstruction &1.6  &2.7 \\
$\Lambda$ reconstruction &1.6 &2.8 \\
Photon selection& 1.5 & 1.5\\
$\mathcal{B}_i^{\rm inter}$& 0.8& 0.8\\
$\chi^{2}_{4\rm C}$ requirement & 0.2 &-        \\
$M_{\mathrm{BC}}$ fit &2.4  &    8.4         \\
MC model &1.4  &    -          \\
Peaking background &  5.0 &1.1\\
Unmatched background& 1.9& 3.7\\
$N_{\Lambda_c^+\bar\Lambda_c^-}$ &1.6&1.4\\
\hline
Total &7.0 &10.7\\
\hline
\hline
\end{tabular}
\end{table}


\begin{itemize}
\item \textit{Tracking and PID}.
 The systematic uncertainties associated with tracking and PID are
investigated‌ using the control samples of $e^+e^-\rightarrow
K^+K^-\pi^+\pi^-$~\cite{BESIII:2019kfh}.  We assign a systematic
uncertainty of 1.0$\%$ for tracking and PID for each proton or charged
pion. Due to the low momentum, the systematic uncertainty of tracking
and PID of a charged kaon is assigned as 2.0\% for each.

\item \textit{The $K_S^0$ reconstruction}.
 The control sample $J/\psi\to K_S^0 K^{\pm}\pi^{\pm}$ is used to
 study the $K_S^0$ reconstruction efficiency~\cite{BESIII:2019kfh}. We
 reweight the detection efficiency of our signal process by the
 data-MC differences in each momentum region. The relative
 differences between the weighted and nominal results are assigned as
 the systematic uncertainties, which are 1.6\% and 2.7\% for
 $\Lambda_c^+\to\Sigma^0K_{S}^{0}\pi^+$ and
 $\Lambda_c^+\to\Sigma^0K_{S}^{0} K^+$, respectively.

\item \textit{The $\Lambda$ reconstruction.}
The $\Lambda$ reconstruction efficiency is studied using the control
sample $J/\psi\to p K^-\bar{\Lambda}$~\cite{BESIII:2018ciw}, and a
similar reweighting procedure as the $K^0_S$ reconstruction is
performed. The
systematic uncertainties are evaluated as the relative difference between
the weighted and nominal efficiencies, which are $1.6\%$ for
$\Lambda_c^+\to\Sigma^0K_{S}^{0}\pi^+$ and $2.8\%$ for
$\Lambda_c^+\to\Sigma^0K_{S}^{0} K^+$.

\item \textit{Photon selection.}  
The uncertainty of the photon reconstruction is studied based on the
control sample $J/\psi\to\pip\pim\piz$ with $\piz\to\gamma\gamma$. For
our signal processes, the energy of the photon is around 0.1 GeV, so we
assign 1.5\% as the systematic uncertainty.

\item \textit{BFs of the intermediate states.} The BFs of intermediate
states ($\Sigma^{0}\to\Lambda\gamma, \Lambda\to p\pim,
K_{S}^{0}\to\pip\pim$) are used as an input in the analysis and quoted
from the PDG~\cite{ParticleDataGroup:2024cfk}; their uncertainties are
propagated, giving a systematic uncertainty of 0.8\%.

 \item \textit{$\chi^{2}_{4\rm C}$ requirement.} The systematic
 uncertainty of the $\chi^{2}_{4\rm C}$ requirement arises from the
 difference between data and MC samples. To study the effect, we fit the
 $\chi^{2}_{4\rm C}$ distribution of data using the MC-simulated
 signal shape convolved with a Gaussian function with free
 parameters. The background shapes are obtained from the inclusive MC
 sample. The Gaussian parameters are used to smear the value of
 $\chi^{2}_{4\rm C}$ of MC sample to obtain new values for the
 efficiency and BF. The difference with the nominal BF is assigned
 as the systematic uncertainty for $\Lambda_c^+\to\Sigma^0K_{S}^{0}
 \pi^+$ process (0.15$\%$). Due to the low statistics of
 $\Lambda_c^+\to\Sigma^0K_{S}^{0} K^+$, we use the same Gaussian
 parameters as $\Lambda_c^+\to\Sigma^0K_{S}^{0}\pip$; the resulting
 difference compared to the nominal BF is found to be negligible.

 \item \textit{$M_{\rm BC}$ fit.} 
The systematic uncertainty of the $M_{\rm BC}$ fit includes those
associated with the signal and background shapes. The uncertainty
associated with the signal MC shape is estimated by changing it to the
Crystal Ball function. The uncertainty due to the background shape of
$\Lambda_c^+\to\Sigma^0K_{S}^{0}\pip$ is estimated with an alternative
background shape obtained from the inclusive MC sample. The total
systematic uncertainty is estimated to be 2.4$\%$. For
$\Lambda_c^+\to\Sigma^{0} K_{S}^{0} K^{+}$, the background shapes are
obtained from inclusive MC samples. Here we change it to the Crystal
Ball function with parameters obtained from the fit to the background
MC sample. The total systematic uncertainty is estimated to be
8.4$\%$.

\item \textit{MC model.} Due to limited statistics, we only use the
PHSP model to generate the signal process. For
$\Lambda_c^+\to\Sigma^{0} K_{S}^{0}\pi^{+}$, we find evidence for the
$\Lambda_c^+\to\Sigma^{0} K^{*+}$ process from the $K_S^{0}\pi^{+}$
mass distribution of data. Therefore, we include this process in the
signal MC generation. The fraction of $\Lambda_c^+\to\Sigma^{0}
K^{*+}$ in the generator is obtained by fitting the
$K_S^{0}\pi^{+}$ mass distribution. The BF difference obtained by
varying the input fraction by $\pm 1\sigma$ is assigned as the
systematic uncertainty (1.4\%). For $\Lambda_c^+\to\Sigma^{0}
K_{S}^{0} K^{+}$, no significant resonance is observed and the MC
simulation models data well. So the systematic uncertainty of the MC
model is ignored.

\item \textit{Peaking background.} 
In the nominal analysis, the yields of the peaking backgrounds are estimated with the efficiency and the BF quoted from the PDG. The input BFs are varied by $\pm1\sigma$ in the fit, and the largest change of the re-measured BF is taken as the systematic uncertainty,
which is 5.0$\%$ for $\Lambda_c^+\to\Sigma^0K_{S}^{0}\pip$ and 1.1$\%$ for $\Lambda_c^+\to\Sigma^0K_{S}^{0} K^+$. 

\item \textit{The unmatched background.} 
To estimate the uncertainty caused by the angle requirement, we vary
it by 5 degrees and obtain a new BF. The BF difference is assigned
as the systematic uncertainty, which is $1.9\%$ for
$\Lambda_c^+\to\Sigma^0K_{S}^{0}\pip$ and $3.7\%$ for
$\Lambda_c^+\to\Sigma^0K_{S}^{0} K^+$.

\item \textit{The number of $N_{\Lambda_c^+\bar{\Lambda}_c^-}$ events.} 
$N_{\Lambda_c^+\bar{\Lambda}_c^-}$ is obtained by the
luminosity~\cite{BESIII:2015qfd,BESIII:2022ulv} and the cross
section~\cite{BESIII:2023rwv}. Its systematic uncertainty is estimated
to be 1.6$\%$ for $\Lambda_c^+\to\Sigma^0K_{S}^{0}\pi^{+}$ and 1.4$\%$
for $\Lambda_c^+\to\Sigma^0K_{S}^{0}K^{+}$.  \end{itemize}

\section{SUMMARY}
\label{sec:summary}
\hspace{1.5em}

By analyzing 6.4 fb$^{-1}$ of $\ee$ annihilation data collected at
center-of-mass energies ranging from $\sqrt{s}=4.600-4.950\gev$ with
the BESIII detector, we report the first observation of $\Lambda_c^+
\to \Sigma^0 K_S^0 \pi^+$ with a statistical significance of
5.9$\sigma$. The measured branching fraction exhibits a notable
difference compared to the theoretical prediction, listed in
Table~\ref{tab:total}, which excludes resonant contributions.
The resonance contribution of $\Lambda^+_c\to
 \Sigma^0K^{*+}$ is evident in Fig.~\ref{fig:KSpi}, albeit with low
 significance. The BF result is
 $\mathcal{B}(\Lambda_c^+ \to \Sigma^0K^{*+}) \times \mathcal{B}(K^{*+} \to
 K^0_S \pi^+) = (0.41\pm0.19\pm0.03)\times 10^{-3}$, by considering the systematic uncertainty similar to $\Lambda_c^+
\to \Sigma^0 K_S^0 \pi^+$.   The resonant contribution
plays a very significant role
 in the $\Lambda_c^+ \to \Sigma^0 K_S^0 \pi^+$ decay.

Furthermore, the BF for the decay channel
$\Lambda_c^+\to\Sigma^0K_{S}^{0} K^+$ has been measured with a
statistical significance of‌ 3.7$\sigma$. We establish an upper limit
at the 90\% C.L. on the branching fraction to be
$\mathcal{B}(\Lambda_c^+\to\Sigma^0K^0_SK^+) < 1.23
\times10^{-3}$. This result is consistent with‌ both
theoretical expectations~\cite{Geng:2024sgq} and the previous BESIII
measurement~\cite{BESIII:2025rda} (Table~\ref{tab:total}). These findings provide crucial
experimental constraints for understanding the decay dynamics and
internal structure of the ground-state charmed baryon $\Lambda_c^{+}$.

\begin{table}[h!]
\centering
\footnotesize
\caption{\label{tab:total} Comparison of branching fractions between experimental results and theoretical predictions which exclude resonant production (in unit of $10^{-3}$). The BF results of this work contain all resonance.}
\begin{tabular}{l|c|c|c} 
\hline
\hline
Decay mode &$\Lambda_c^+\to\Sigma^0K_S^{0}\pip$   &$\Lambda_c^+ \to \Sigma^0 K^{*+} (K^{*+}\to K^0_S\pi^+)$& $\Lambda_c^+\to\Sigma^0K_S^{0}K^{+}$ \\ 
\hline
Theory calculations&$(0.17\pm0.05)$~\cite{Geng:2024sgq}& $(0.40\pm0.10)$~\cite{Hsiao:2019yur}&$(0.12\pm0.04)$~\cite{Geng:2024sgq} \\
\hline
Experimental results &-&-&$<1.28$~\cite{BESIII:2025rda}\\
\hline
This work& $(0.58 \pm 0.14_{\rm stat.} \pm 0.04_{\rm syst.})$ &$(0.41\pm0.19\pm0.03)$&$(0.35\pm0.16_{\rm stat.}\pm0.04_{\rm syst.})$\\
&&&$<1.23$\\
\hline
\hline
\end{tabular}
\end{table}

\newpage
\acknowledgments
\hspace{1.5em}

We thank Y. K. Hsiao for useful discussions.
The BESIII Collaboration thanks the staff of BEPCII (https://cstr.cn/31109.02.BEPC) and the IHEP computing center for their strong support. This work is supported in part by National Key R\&D Program of China under Contracts Nos. 2023YFA1606000, 2023YFA1606704, 2023YFA1609400; National Natural Science Foundation of China (NSFC) under Contracts Nos. 12305105, 12205141, 12105127, 11635010, 11935015, 11935016, 11935018, 12025502, 12035009, 12035013, 12061131003, 12192260, 12192261, 12192262, 12192263, 12192264, 12192265, 12221005, 12225509, 12235017, 12342502, 12361141819; the Chinese Academy of Sciences (CAS) Large-Scale Scientific Facility Program; The Strategic Priority Research Program of Chinese Academy of Sciences under Contract No. XDA0480600; CAS under Contract No. YSBR-101; Natural Science Foundation of Shandong Province under Grants No. ZR2023QA119; 100 Talents Program of CAS; The Institute of Nuclear and Particle Physics (INPAC) and Shanghai Key Laboratory for Particle Physics and Cosmology; ERC under Contract No. 758462; German Research Foundation DFG under Contract No. FOR5327; Istituto Nazionale di Fisica Nucleare, Italy; Knut and Alice Wallenberg Foundation under Contracts Nos. 2021.0174, 2021.0299, 2023.0315; Ministry of Development of Turkey under Contract No. DPT2006K-120470; National Research Foundation of Korea under Contract No. NRF-2022R1A2C1092335; National Science and Technology fund of Mongolia; Polish National Science Centre under Contract No. 2024/53/B/ST2/00975; STFC (United Kingdom); Swedish Research Council under Contract No. 2019.04595; U. S. Department of Energy under Contract No. DE-FG02-05ER41374

\newpage
M.~Ablikim$^{1}$\BESIIIorcid{0000-0002-3935-619X},
M.~N.~Achasov$^{4,d}$\BESIIIorcid{0000-0002-9400-8622},
P.~Adlarson$^{81}$\BESIIIorcid{0000-0001-6280-3851},
X.~C.~Ai$^{87}$\BESIIIorcid{0000-0003-3856-2415},
C.~S.~Akondi$^{31A,31B}$\BESIIIorcid{0000-0001-6303-5217},
R.~Aliberti$^{39}$\BESIIIorcid{0000-0003-3500-4012},
A.~Amoroso$^{80A,80C}$\BESIIIorcid{0000-0002-3095-8610},
Q.~An$^{77,64,\dagger}$,
Y.~H.~An$^{87}$\BESIIIorcid{0009-0008-3419-0849},
Y.~Bai$^{62}$\BESIIIorcid{0000-0001-6593-5665}
% ...
\\
\vspace{0.2cm}
(BESIII Collaboration)\\
\vspace{0.2cm} {\it
$^{1}$ Institute of High Energy Physics, Beijing 100049, People's Republic of China\\
% ...
$^{4}$ Budker Institute of Nuclear Physics SB RAS (BINP), Novosibirsk 630090, Russia\\
% ...
$^{31}$ INFN Sezione di Ferrara, (A)INFN Sezione di Ferrara, I-44122, Ferrara, Italy; (B)University of Ferrara, I-44122, Ferrara, Italy\\
% ...
$^{39}$ Johannes Gutenberg University of Mainz, Johann-Joachim-Becher-Weg 45, D-55099 Mainz, Germany\\
% ...
$^{62}$ Southeast University, Nanjing 211100, People's Republic of China\\
% ...
$^{64}$ State Key Laboratory of Particle Detection and Electronics, Beijing 100049, Hefei 230026, People's Republic of China\\
% ...
$^{77}$ University of Science and Technology of China, Hefei 230026, People's Republic of China\\
% ...
$^{80}$ University of Turin and INFN, (A)University of Turin, I-10125, Turin, Italy; (B)University of Eastern Piedmont, I-15121, Alessandria, Italy; (C)INFN, I-10125, Turin, Italy\\
$^{81}$ Uppsala University, Box 516, SE-75120 Uppsala, Sweden\\
% ...
$^{87}$ Zhengzhou University, Zhengzhou 450001, People's Republic of China\\

\vspace{0.2cm}
$^{\dagger}$ Deceased\\
% ...
$^{d}$ Also at the Novosibirsk State University, Novosibirsk, 630090, Russia\\
% ...

}


\begin{thebibliography}{99}
\bibitem{Politzer:1973QCD1}
H.~D.~Politzer, 
\href{https://doi.org/10.1103/PhysRevLett.30.1346}
{Phys. Rev. Lett. \textbf{30}, 1346-1349 (1973)}. 
\bibitem{Gross:1973QCD2}
D.~J.~Gross and F.~Wilczek,
\href{https://doi.org/10.1103/PhysRevLett.30.1343}
{Phys. Rev. Lett. \textbf{30}, 1343-1346 (1973)}.


\bibitem{BESIII:2020LambdacSP}
M.~Ablikim \textit{et al.} (BESIII Collaboration),
\href{https://doi.org/10.1103/PhysRevD.103.L091101}
{Phys. Rev. D \textbf{103}, L091101 (2021)}.


\bibitem{Abrams:1979FirstLambdac}
G.~S.~Abrams, M.~S.~Alam, C.~A.~Blocker, A.~Boyarski, M.~Breidenbach, D.~L.~Burke, W.~C.~Carithers, W.~Chinowsky, M.~W.~Coles and S.~Cooper, \textit{et al.}
\href{https://doi.org/10.1103/PhysRevLett.44.10}
{Phys. Rev. Lett. \textbf{44}, 10 (1980)}.




\bibitem{Cheng:2015iom}
H.~Y.~Cheng,
\href{https://doi.org/10.1007/s11467-015-0483-z}
{Front. Phys. (Beijing) \textbf{10}, 101406 (2015)}.
\bibitem{Cheng:2021qpd}
H.~Y.~Cheng,
\href{https://doi.org/10.1016/j.cjph.2022.06.021}
{Chin. J. Phys. \textbf{78}, 324-362 (2022)}.
\bibitem{Li:2021iwf}
H.~B.~Li and X.~R.~Lyu,
\href{https://doi.org/10.1093/nsr/nwab181}
{Natl. Sci. Rev. \textbf{8}, nwab181 (2021)}.


\bibitem{Li:2025nzx}
P.~R.~Li, X.~R.~Lyu and Y.~Zheng,
\href{https://iopscience.iop.org/article/10.1088/1674-1137/ae1187}{Chin. Phys. \textbf{50}, 022002 (2026).}


\bibitem{Wang:2024wrm}
H.~J.~Wang, P.~R.~Li, X.~R.~Lyu, J.~Tandean and H.~B.~Li,
\href{https://www.sciencedirect.com/science/article/pii/S2095927325001963?via\%3Dihub}{Sci. Bull. \textbf{70}, 1183-1185 (2025).}


\bibitem{Yu:2017zst}
F.~S.~Yu, H.~Y.~Jiang, R.~H.~Li, C.~D.~L\"u, W.~Wang and Z.~X.~Zhao,
\href{https://iopscience.iop.org/article/10.1088/1674-1137/42/5/051001}{Chin. Phys. C \textbf{42}, 051001 (2018)}.

\bibitem{Uppal:1994pt}
T.~Uppal, R.~C.~Verma and M.~P.~Khanna,
\href{https://journals.aps.org/prd/abstract/10.1103/PhysRevD.49.3417}{Phys. Rev. D \textbf{49}, 3417-3425 (1994)}.

\bibitem{Tetlalmatzi-Xolocotzi:2024ztd}
G.~Tetlalmatzi-Xolocotzi,
\href{https://inspirehep.net/files/0456b6ce354f55d9e594b6ceb27030dd}{PoS \textbf{DISCRETE2022}, 040 (2024).}



\bibitem{Zenczykowski:1993hw}
P.~Zenczykowski,
\href{https://journals.aps.org/prd/abstract/10.1103/PhysRevD.50.402}{Phys. Rev. D \textbf{50}, 402-411 (1994)}.


\bibitem{Cheng:1991sn}
H.~Y.~Cheng and B.~Tseng,
\href{https://journals.aps.org/prd/pdf/10.1103/PhysRevD.46.1042}{Phys. Rev. D \textbf{46}, 1042 (1992)};
\href{https://journals.aps.org/prd/abstract/10.1103/PhysRevD.55.1697}{[erratum: Phys. Rev. D \textbf{55}, 1697 (1997)].}

\bibitem{Cheng:2018hwl}
H.~Y.~Cheng, X.~W.~Kang and F.~Xu,
\href{https://journals.aps.org/prd/abstract/10.1103/PhysRevD.97.074028}{Phys. Rev. D \textbf{97}, 074028 (2018)}.


\bibitem{Savage:1989qr}
M.~J.~Savage and R.~P.~Springer,
\href{https://journals.aps.org/prd/abstract/10.1103/PhysRevD.42.1527}{Phys. Rev. D \textbf{42}, 1527-1543 (1990)}.


\bibitem{Geng:2017mxn}
C.~Q.~Geng, Y.~K.~Hsiao, C.~W.~Liu and T.~H.~Tsai,
\href{https://link.springer.com/article/10.1007/JHEP11(2017)147}{JHEP \textbf{11}, 147 (2017)}.


\bibitem{Geng:2024sgq}
C.~Q.~Geng, C.~W.~Liu and S.~L.~Liu,
\href{https://journals.aps.org/prd/abstract/10.1103/PhysRevD.109.093002}{Phys. Rev. D \textbf{109}, 093002 (2024)}.

\bibitem{ParticleDataGroup:2024cfk}
S.~Navas \textit{et al.} (Particle Data Group),
\href{https://journals.aps.org/prd/abstract/10.1103/PhysRevD.110.030001}{Phys. Rev. D \textbf{110}, 030001 (2024)}.





\bibitem{BESIII:2025rda}
M.~Ablikim \textit{et al.} (BESIII Collaboration),
\href{https://journals.aps.org/prd/abstract/10.1103/csfm-p3h6}{Phys. Rev. D \textbf{112}, 032006 (2025).}


\bibitem{Geng:2018upx}
C.~Q.~Geng, Y.~K.~Hsiao, C.~W.~Liu and T.~H.~Tsai,
\href{https://journals.aps.org/prd/abstract/10.1103/PhysRevD.99.073003}{Phys. Rev. D \textbf{99}, 073003 (2019).}


\bibitem{Wang:2025uie}
Y.~L.~Wang and Y.~K.~Hsiao,
\href{https://arxiv.org/pdf/2505.21311}{[arXiv:2505.21311 [hep-ex]]}.

\bibitem{BESIII:2024mbf}
M.~Ablikim \textit{et al.} (BESIII Collaboration),
\href{https://journals.aps.org/prl/pdf/10.1103/PhysRevLett.134.021901}{Phys. Rev. Lett. \textbf{134}, 021901 (2025)}. 


\bibitem{BESIII:2024xny}
M.~Ablikim \textit{et al.} (BESIII Collaboration),
\href{https://journals.aps.org/prd/pdf/10.1103/PhysRevD.111.012014}{Phys. Rev. D \textbf{111}, 012014 (2025)}.

\bibitem{Hsiao:2019yur}
Y.~K.~Hsiao, Y.~Yao and H.~J.~Zhao,
\href{https://doi.org/10.1016/j.physletb.2019.03.031}{Phys. Lett. B \textbf{792}, 35-39 (2019)}.






\bibitem{Geng:2020zgr}
C.~Q.~Geng, C.~W.~Liu and T.~H.~Tsai,
\href{https://journals.aps.org/prd/pdf/10.1103/PhysRevD.101.053002}{Phys. Rev. D \textbf{101}, 053002 (2020)}.


\bibitem{BESIII:2022ulv}
M.~Ablikim \textit{et al.} (BESIII Collaboration),
\href{https://iopscience.iop.org/article/10.1088/1674-1137/ac84cc}{Chin. Phys. C \textbf{46}, 113003 (2022)}.









\bibitem{Ablikim:2009aa}
  M.~Ablikim {\it et al.} (BESIII Collaboration),
 \href{https://doi.org/10.1016/j.nima.2009.12.050}{Nucl.\ Instrum.\ Meth.\ A {\bf 614}, 345 (2010).}

\bibitem{Yu:IPAC2016-TUYA01}
   C.~H.~Yu {\it et al.},
  Proceedings of IPAC2016, Busan, Korea, 2016,
  \href{https://inspirehep.net/files/96083dc6a03caf20597ac55b4500673a}{doi:10.18429/JACoW-IPAC2016-TUYA01.}
  
  
  \bibitem{Ablikim:2019hff}
  M.~Ablikim {\it et al.} (BESIII Collaboration),
  \href{https://iopscience.iop.org/article/10.1088/1674-1137/44/4/040001)}{Chin. Phys. C {\bf 44}, 040001 (2020).}

\bibitem{etof}
 X.~Li {\it et al.}, \href{https://link.springer.com/article/10.1007/s41605-017-0014-2}{Radiat. Detect. Technol. Methods {\bf 1}, 13 (2017);}
 Y.~X.~Guo {\it et al.}, \href{https://link.springer.com/article/10.1007/s41605-017-0012-4}{Radiat. Detect. Technol. Methods {\bf 1}, 15 (2017);}
 P.~Cao {\it et al.}, \href{https://linkinghub.elsevier.com/retrieve/pii/S0168900219314068}{Nucl.\ Instrum.\ Meth.\ A {\bf 953}, 163053 (2020).}



\bibitem{geant4}
  S.~Agostinelli {\it et al.} (GEANT4 Collaboration),
  \href{https://linkinghub.elsevier.com/retrieve/pii/S0168900203013688}{Nucl.\ Instrum.\ Meth.\ A {\bf 506}, 250 (2003).}

\bibitem{ref:kkmc}
  S.~Jadach, B.~F.~L.~Ward and Z.~Was,
  \href{https://journals.aps.org/prd/abstract/10.1103/PhysRevD.63.113009}{Phys.\ Rev.\ D {\bf 63}, 113009 (2001);}
  \href{https://linkinghub.elsevier.com/retrieve/pii/S0010465500000485}{Comput.\ Phys.\ Commun.\  {\bf 130}, 260 (2000).}

\bibitem{ref:evtgen}
  D.~J.~Lange,
  \href{https://linkinghub.elsevier.com/retrieve/pii/S0168900201000894}{Nucl.\ Instrum.\ Meth.\ A {\bf 462}, 152 (2001);}
  R.~G.~Ping,
  \href{https://iopscience.iop.org/article/10.1088/1674-1137/32/8/001}{Chin. Phys. C {\bf 32}, 599 (2008).}




\bibitem{ref:lundcharm}
  J.~C.~Chen, G.~S.~Huang, X.~R.~Qi, D.~H.~Zhang and Y.~S.~Zhu,
  \href{https://journals.aps.org/prd/abstract/10.1103/PhysRevD.62.034003}{Phys.\ Rev.\ D {\bf 62}, 034003 (2000);}
  R.~L.~Yang, R.~G.~Ping and H.~Chen,
  \href{https://iopscience.iop.org/article/10.1088/0256-307X/31/6/061301}{Chin.\ Phys.\ Lett.\  {\bf 31}, 061301 (2014).}



\bibitem{photos2}
E.~Barberio, B.~van Eijk and Z.~Was,
\href{https://linkinghub.elsevier.com/retrieve/pii/001046559190012A}{Comput. Phys. Commun. {\bf 66}, 115 (1991).}

\bibitem{BESIII:2023rwv}
M.~Ablikim \textit{et al.} (BESIII Collaboration),
\href{https://journals.aps.org/prl/abstract/10.1103/PhysRevLett.131.191901}{Phys. Rev. Lett. \textbf{131}, 191901 (2023).}





\bibitem{BESIII:2015qfd}
M.~Ablikim \textit{et al.} (BESIII Collaboration),
\href{https://iopscience.iop.org/article/10.1088/1674-1137/39/9/093001}{Chin. Phys. C \textbf{39}, 093001 (2015).}

\bibitem{ARGUS:1990hfq}
H.~Albrecht \textit{et al.} (ARGUS Collaboration),
\href{https://doi.org/10.1016/0370-2693(90)91293-K}{Phys. Lett. B \textbf{241}, 278-282 (1990).}

\bibitem{BESIII:2019kfh}
M.~Ablikim \textit{et al.} (BESIII Collaboration),
\href{https://journals.aps.org/prd/abstract/10.1103/PhysRevD.99.112005}{Phys. Rev. D \textbf{99}, 112005 (2019)}.

\bibitem{BESIII:2018ciw}
M.~Ablikim \textit{et al.} (BESIII Collaboration),
\href{https://journals.aps.org/prl/abstract/10.1103/PhysRevLett.121.062003}{Phys. Rev. Lett. \textbf{121}, 062003 (2018).}



\end{thebibliography}
\end{document}